# They'll Know It When They See It:
# Analyzing Post-Release Feedback from the Android Community

Sherlock A. Licorish, Chan Won Lee, Bastin Tony Roy Savarimuthu, Priyanka Patel and
Stephen G. MacDonell

*Department of Information Science*
*University of Otago*
*PO Box 56, Dunedin 9016, New Zealand*
*sherlock.licorish@otago.ac.nz, leech041@student.otago.ac.nz, tony.savarimuthu@otago.ac.nz,*
*priyanka.patel@otago.ac.nz, stephen.macdonell@otago.ac.nz*

## Abstract

*It is known that user involvement and user-centered design enhance system acceptance, particularly when end-users' views are considered early in the process. However, the increasingly common method of system deployment, through frequent releases via an online application distribution platform, relies more on post-release feedback from a virtual community. Such feedback may be received from large and diverse communities of users, posing challenges to developers in terms of extracting and identifying the most pressing requests to address. In seeking to tackle these challenges we have used natural language processing techniques to study enhancement requests logged by the Android community. We observe that features associated with a specific subset of topics were most frequently requested for improvement, and that end-users expressed particular discontent with the Jellybean release. End-users also tended to request improvements to specific issues together, potentially posing a prioritization challenge to Google.*

**Keywords**: *Virtual Communities, Online Application Distribution, Android, Enhancement Requests, Post-release Feedback, Natural Language Processing*

## 1. Introduction

The benefits of user involvement and user-centered design to those involved in systems development and deployment have been long established. For instance, previous work has determined that software projects are most successful when end-users' criticisms are accommodated (McKeen et al. 1994), and others have linked the acceptance of end-users' feedback to their satisfaction with or acceptance of a product (Lin and Shao 2000). Thus, consideration of end-user feedback relates directly to the satisfaction of their needs and expectations.

With growing interest in virtual communities and online application distribution platforms (OADPs, e.g., Apple AppStore[1] and Google Play[2]), the means of satisfying end-users' needs and expectations has changed. In fact, end-users' needs may not be explicitly considered prior to software release. Instead, developers typically identify a possible target market for an app[3], develop and release that product, and then perfect it after use by the community, challenging the assumptions of previous user-focused models of development. Those models (e.g., in-house or bespoke development) generally required a sample of potential users of the system to perform early user acceptance testing (UAT) before a beta version of the software was released, having benefited from this round of preliminary user evaluation and thus limiting discontent. In contrast, on OADPs apps are typically released to an entire community of users prior to them knowing about the plan to deliver them. That said, the desire to satisfy the *potential* needs of users, and so encourage uptake, remains a priority. Thus, post-release feedback has become integral to development in driving corrective and perfective efforts (resolving bugs, and enhancing features and adding new functionality, respectively) in OADPs. Many users in these settings are happy to 'vent their feelings', whether to express satisfaction or dissatisfaction after using apps (Ko et al. 2011). In some cases users are also able to rate products, informing other potential users. Applications may also appear on 'top lists' (e.g., for downloads, costs, user ratings), which also influences other users' decisions about adopting them. Users might also describe their experience using apps on other fora, including social networks (Yamakami 2011).

---

[1] https://itunes.apple.com/en/genre/ios/id36?mt=8

[2] https://play.google.com/store/apps?hl=en

[3] We use the term 'app' to describe a program or software product that is frequently delivered (especially to mobile devices) via an online platform

In addition to improving the quality of previously delivered features, such community feedback also helps to identify features (i.e., new requirements) desired by end-users, and so informs apps' evolution. One of the challenges faced by developers in understanding this feedback, however, is the labor-intensive nature of extracting insights from end-user reviews. This is particularly the case when user communities are large. Thus, efforts aimed at augmenting manual extraction of knowledge are pertinent to supporting delivery of the 'right' features to this virtual community. Previous work has indeed demonstrated the utility of such efforts, employing complex aspect identification and sentiment analysis techniques in the delivery of useful insights, e.g., Yu et al. (2011). In the same vein, other deeper contextual analysis approaches have also been useful, albeit focusing on a smaller subset of issues (Licorish et al. 2015).

One such virtual community has formed around the Android operating system (OS), the most widely used mobile OS (Butler 2011). Investigating the recorded views of this community would have two particular benefits: one, it would be useful in terms of understanding the past and possible future of the OS; and two, the utility of particular analysis methods could be evaluated in this context. We thus seek to identify the features users were most dissatisfied with and the issues developers should target. We acknowledge that the Android OS community might not be thought of as a typical OADP, where developers commonly consist of individuals or small teams that have limited resources and may not be able to conduct extensive tests of their product before launch. Instead, the Android OS is sponsored primarily by Google, whose significant resources may be employed for UAT, and thus, this product may be of very good quality. However, given that the Android issue tracker brings together developers *and* users, such that users can provide diverse feedback on this product's performance and improvement opportunities, we believe that lessons learned from such a case may be relevant to OADPs.

In the next section we present the study's background and motivation, and we outline our specific research hypotheses. We then describe our research setting, introducing our measures in this section. Thereafter, we present our results, before discussing our findings and outlining their implications. Finally, threats to the validity of our study are outlined.

## 2. Background and Motivation

### *The Relevance of End-users' Feedback*

End-users have long played an integral role in software product development and in determining whether delivered systems benefit from prolonged use (McKeen et al. 1994). These 'community members' interact with software systems to unearth bugs and to identify new features and improvement opportunities. Such activities are generally considered under the established sub-disciplines at the intersection of behavior and information technology: user involvement, usability and human-computer-interaction (Kujala 2003).

The consensus in these sub-disciplines is that involving users during what was traditionally known as the system development life cycle, particularly in early stages, provides critical insights into the likely context of system use. Awareness of this context then leads to the development and delivery of systems that are appreciated by users. There are multiple means of achieving this awareness. For instance, developers may gather design insights by observing users performing specific activities, with a goal of then providing solutions to aid such processes. Developers may also review artefacts and documents created and utilized by users, incorporating them into software interfaces. Furthermore, developers may observe users working with simulations and prototypes they provide (Gould and Lewis 1985). These activities then inform improvements in the quality of the system that is delivered along with reduced likelihood of developers creating unwanted system features, leading to an increase in overall user acceptance of the software system (Kujala 2003). Newer models of software development (and OADP in particular), however, do not adhere to such a process; we consider this issue next.

### *Accommodating End-users' Feedback in OADPs*

As just noted, formal mechanisms involving pre-emptive user involvement and user-centered design are associated with traditional software development contexts, wherein developers and users can interact in a shared physical space. Clearly such approaches are not exploited in virtual communities and OADPs. This presents challenges for the app community in terms of ensuring the delivery of high-quality solutions; sometimes even apps that are rated highly possess many avenues for improvement (Pagano and Maalej 2013). On the other hand, this new context affords an alternative opportunity to implicitly capture a wide range of user views and desires from their comments and reviews, not previously possible when considering a small cohort of potential system users.

A growing number of studies have therefore focused on using such sources to identify improvement opportunities. For instance, in their analysis of 3,279 reviews of 161 apps on Google Play, researchers observed that over 23% of users' comments were feature requests, and apps that were rated more highly received more requests than their lower-rated counterparts (Iacob and Harrison 2013). An experiment involving over one million reviews from 1,100 apps on the Apple AppStore revealed that positive reviews were correlated with higher ratings, and negative feedback provided little information for app improvement (Pagano and Maalej 2013).

One extremely large community that has not received such attention is that associated with the Android OS. While the Android OS issue tracker has been the subject of research, previous work has largely considered the localization of bugs based on issues that were logged. For instance, Kumar Maji et al. (2010) studied 758 bugs from four early versions of the system (versions 1.1, 1.5, 1.6 and 2.0) and found most bugs to be present in the application layer. Bhattacharya et al. (2013) found that, while Android security bugs are recorded with higher quality (e.g., fuller description, reproducible steps and output details), their mean resolution time was greater than that of other bugs. Beyond bug-related issues, it would be useful to understand the features most sought after by Android end-users', in support of both this virtual community and the developers who serve it. We thus used this case to test three hypotheses, as outlined next.

### *Hypotheses Development*

While previous work has considered where most software issues are localized (Kumar Maji et al. 2010), the nature and volume of these issues have not been considered. Applications of the Pareto principle to software development have shown that power-law behavior applies to software fault localization – 80% of software defects are typically found in 20% of the components (e.g., Boehm 1987) – suggesting that specific types of features consume most developer rework effort (Louridas et al. 2008). It would be useful to evaluate if such principles hold across environments, and especially considering OADPs, where such insights could be key to retaining expertise that built specific features to address emerging end-users' concerns. We thus designed hypothesis one (H1) to address this issue.

> *H1. Specific Android features are most often the subjects of enhancement requests.*

There has been divergence in the literature around the likely impact of frequent releases, and the scope of each release, on the quality of software development outcomes. This issue is central to OADPs, where the goal is to satisfy end-users after software applications are delivered, as against prior to delivery in conventional environments. While some have argued that fewer bugs appear in some releases than in others, particularly when considering the length of the release cycle (Baysal et al. 2011), others have observed the opposite (Khomh et al. 2012). Another thread of research has established that the need for enhancements decreases as a software project matures (Hindle et al. 2011), as against variance in release frequency. Thus, the development of a generalizable theory for this phenomenon has been out of researchers' reach. We propose to add to this body of work by testing hypothesis two (H2).

> *H2. Specific Android releases will have more requests for enhancement.*

Complex network theory notes that social and biological systems display interconnections that are unique, but not necessarily regular or random (Estrada 2011). Although still developing, this theory has been applied to the study of software developers' dialogues and the artefacts they produce, with results confirming the usefulness of its application. For instance, studies have shown that analyzing developers' networks improves bug prediction models (Zimmermann and Nagappan 2008). Complex network theory has also been applied to the study of bug reporters' details to inform bug triaging procedures (Zanetti et al. 2013), and for understanding the effects of requirement dependency on early software integration bugs (Junjie et al. 2013), providing promising outcomes. Given the way software components are themselves interconnected, using complex network theory as a basis for studying the relationship between issues reported in enhancement requests could help us to uncover complex webs of defects reported by end-users in OADPs. We thus outline our third hypothesis (H3) to address this issue.

> *H3. Enhancement requests are likely to take issue with multiple features.*

## 3. Research Setting

We used the Android OS community as our case study 'organization'. Issues identified by the Android community are submitted to an issue tracker hosted by Google (http://tinyurl.com/njmeh4). Among the data stored in the issue tracker are the following details: Issue ID, Type, Summary description, Stars (number of people following the issue), Open date, Reporter, Reporter Role, and OS Version. We extracted a snapshot of the issue tracker, comprising issues submitted between January 2008 and March 2014. Our particular snapshot comprised 21,547 issues. These issues were imported into a Microsoft

SQL database, and thereafter, we performed data cleaning by executing previously written scripts to remove all HTML tags and foreign characters (Licorish and MacDonell 2013), and particularly those in the summary description field, to avoid these confounding our analysis.

We next employed exploratory data analysis (EDA) techniques to investigate the data properties, conduct anomaly detection and select our research sample. Issues were labelled (by their submitters) as defects (15,750 issues), enhancements (5,354 issues) and others (438 issues). Given our goal of studying the desired improvements of the Android community we selected the 5,354 enhancement requests, as logged by 4,001 different reporters. Of the 5,354 enhancement requests, 577 were logged by members identifying themselves as developers, 328 were sourced from users, and 4,449 were labelled as anonymous. We examined the data of each request in our database to align these with the commercial releases of the Android OS. Its first release was in September 2008 (http://tinyurl.com/pr72od4), although the first issue was logged in the issue tracker in January 2008. This suggests that the community was already actively engaged with the Android OS after the release of the first beta version in November 2007 (http://tinyurl.com/m29nmow), with issues being reported just two months later. Given this level of active engagement, occurring even before the first official Android OS release, we partitioned the issues based on Android OS release date and major name change. So for instance, all of the issues logged from January 2008 (the date the first issue was logged) to February 2009 (the date of an Android release before a major name change) were labelled 'Early versions', reflecting the period occupied by Android OS releases 1.0 and 1.1 which were both without formal names. The subsequent partition comprised the period between Android OS version 1.1 and Cupcake (Android version 1.5), and so on.

Table 1 provides a brief summary of the numbers of enhancement requests logged between each of the major releases, from the very first release through to KitKat – Android version 4.4. From column three of Table 1 (Number of days between releases) it can be noted that the time taken between the delivery of most of Android OS's major releases (those involving a name change) was between 80 and 156 days, with two official releases (Gingerbread and Jellybean) falling outside this range. The fourth column of Table 1 (Total requests logged) shows that the number of enhancement requests reported increased somewhat as the Android OS progressed, with this rise being particularly evident when the mean requests reported per day for each release is considered (refer to the values in the fifth column for details). Over the six years of Android OS's existence, on average, 2.7 enhancement requests were logged every day (median = 2.6, Std Dev = 2.1).

| Version (Release) | Last release date | Number of days between releases | Total requests logged | Mean requests per day |
|---|---|---|---|---|
| Early versions (1.0, 1.1) | 09/02/2009 | 451 | 173* | 0.4 |
| Cupcake (1.5) | 30/04/2009 | 80 | 64 | 0.8 |
| Donut (1.6) | 15/09/2009 | 138 | 141 | 1.0 |
| Éclair (2.0, 2.01, 2.1) | 12/01/2010 | 119 | 327 | 2.8 |
| Froyo (2.2) | 20/05/2010 | 128 | 349 | 2.7 |
| Gingerbread (2.3, 2.37) | 09/02/2011 | 265 | 875 | 3.3 |
| Honeycomb (3.0, 3.1, 3.2) | 15/07/2011 | 156 | 372 | 2.4 |
| Ice Cream Sandwich (4.0, 4.03) | 16/12/2011 | 154 | 350 | 2.3 |
| Jellybean (4.1, 4.2, 4.3) | 24/07/2013 | 586 | 1,922 | 3.3 |
| KitKat (4.4) | 31/10/2013 | 99 | 781 | 7.9 |
| | | $\sum$ = 2,176 | $\sum$ = 5,354 | $\bar{x}$ = 2.7 |

* Total number of requests logged between the first beta release on 16/11/2007 and Android version 1.1 released on 09/02/2009

**Table 1. Android OS enhancement requests over the major releases**

We employed natural language processing (NLP) techniques to study these enhancement requests in evaluating our three hypotheses. NLP techniques are often used to explore and understand language usage within groups and societies. We employed multiple techniques from the NLP space in our analysis, including corpus linguistic part-of-speech tagging (POS) (Toutanova et al. 2003), computational linguistic n-gram analysis (Manning and Schtze 1991), and pointwise mutual information (PMI) measurement (Church and Hanks 1990). We then conducted multiple reliability checks. These methods are now described in turn.

### NLP Techniques

Prior research has established that noun terms in unstructured text reflect the main concepts in the subject of a clause. From a POS perspective, nouns are indeed reflective of specific objects or things (http://tinyurl.com/nvh3kf). From a linguistic perspective, nouns often form the subjects and objects of clauses or verb phrases (http://tinyurl.com/ovgu2zd). These and other understandings have been embedded as rules in NLP tools, including the Stanford parser which performs POS tagging (Toutanova et al. 2003). We created a program that incorporated the Stanford API to enable us to extract noun phrases from the enhancement requests, before counting the frequency of each noun as unigrams (e.g., if "SMS" appeared at least once in 20 enhancement requests our program would output SMS = 20). This exercise allowed us to investigate *whether specific features were most often the subjects of enhancement requests*. The ranking of words in this manner draws from computational linguistics, and is referred to as n-gram analysis. The n-gram is defined as a continuous sequence of n words (or characters) in length that is extracted from a larger stream of elements (Manning and Schtze 1991). We also investigated *whether specific Android releases had more requests for enhancement* by executing the above mentioned program against enhancement requests logged across each release block in Table 1. Thereafter, we investigated *whether enhancement requests took issue with multiple features* using PMI, which measures the degree of statistical dependence between two words. Our implementation of PMI considers the syntactic relations between features in each request, by providing noun-pair similarity measures. We formally define this approach as: PMI $(n_1, n_2) = \log_2 (p(n_1, n_2)/p(n_1)p(n_2))$, where $p(n_1, n_2)$ represents the probability that $noun_1$ co-occurs with $noun_2$, and $p(n_1)$ and $p(n_2)$ are the marginal probabilities of the occurrence of $noun_1$ and $noun_2$ (Church and Hanks 1990).

### *Reliability Assessments*

The first two authors of this work triangulated the NLP findings by manually coding a random sample of 50 outputs from the POS, n-gram and PMI analyses, to check that nouns were correctly classified, and to verify the co-occurrence of requests (flagging each as true or false). We computed reliability measurements from these coding outputs using Holsti's coefficient of reliability (Holsti 1969) to evaluate our agreement. For our first reliability check (that nouns were correctly classified) we observed a 90% agreement, and the remaining 10% of codes were resolved by consensus. We observed 100% agreement for our second observation (where aspects were reported in conjunction).

## 4. Results

### *Features that are Most Often the Subjects of Enhancement Requests (H1)*

Given space limitations we report here the 20 features (of 316) that the Android community mentioned most often across their enhancement requests (as shown in Figure 1). In rank order the Android community identified contacts (320 requests), screen (243 requests), notification (232 requests), call (202 requests), calendar (172 requests), email (166 requests), text (163 requests) and keyboard (152 requests) as most often needing enhancement.

Requests regarding these eight features were logged more than the average count of 147.6 (median = 136.5, Std Dev = 62.4) for the top 20 frequently occurring enhancement requests. Just below this average were requests for enhancement related to button (144 requests), api[4] (139 requests) and sms (134 requests). There were slightly fewer requests for enhancement for number (124 requests), message (117 requests), file (107 requests), browser (95 requests) and volume (95 requests); while default settings (89 requests), alarm (87 requests), voice (87 requests) and search (84 requests) had fewer requests still.

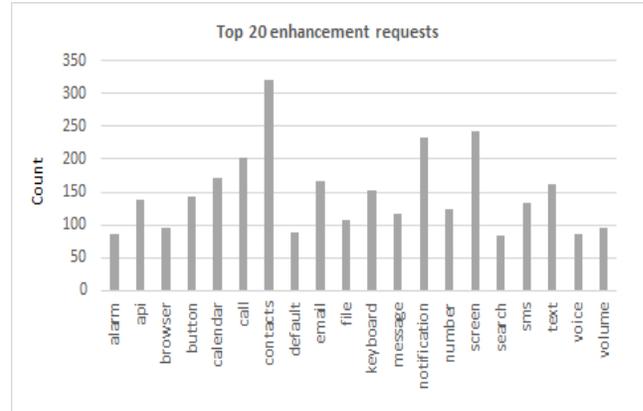

**Figure 1. Features mentioned in Android's top 20 most frequent enhancement requests**

We investigated the relative attention these issues received by examining the number of stars (i.e., followers) the top 20 features attracted. This result is shown in Figure 2, which reveals a slightly different pattern to that shown in Figure 1. We observe that, on average, more individuals followed api (80.5 followers per request), calendar (60.9 followers each), file (36 followers each), text (24.7 followers each), voice (24.5 followers each) and notification (22.6 followers each) requests. These requests attracted more than the average 20.4 followers that were monitoring the top 20 feature enhancement requests (median = 14.1, Std Dev = 20.1). Of these enhancement requests, notification, calendar and keyboard were reported more than usual (see above), while api requests were just below the average. File and voice requests were far fewer, although these also received above average numbers of followers. Figure 2 shows that message, alarm, button, screen, browser, keyboard and email requests were tracked by fewer than half the mean number of followers (i.e., < 10 followers). While we did not observe an 80%-20% split for requests above the average count of 147.6 and others (we noted a 56%-44% division), a Kendall's tau-b non-parametric correlation test confirmed that features that were mentioned more frequently across enhancement

---

[4] This feature groups various Android api requests (e.g., file storage api, camera api, and so on).

requests also received greater levels of community attention (τ = 0.40, *P* < 0.05).

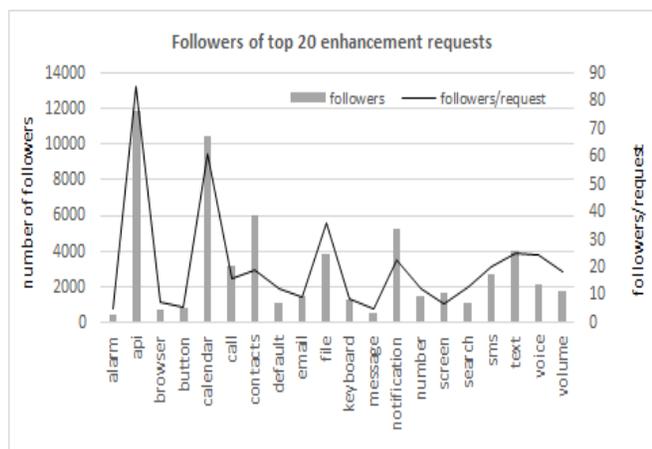

**Figure 2. Followers of features mentioned in Android's top 20 most frequent enhancement requests**

## Android Releases and Requests for Enhancement (H2)

For the top eight feature (i.e., contacts, screen, notification, call, calendar, email, text and keyboard), Figure 3 shows that relatively few requests were made after the initial releases, but this trend changed after the releases of Éclair, when the Android community sought increased numbers of enhancements in all of the eight features considered here. Apart from notification enhancement requests, the desire for enhancement continued after the release of Froyo. After the release of Gingerbread the community appears to have been particularly dissatisfied, requesting the highest numbers of enhancements up to this point (notwithstanding the potentially growing user base). This pattern was reversed after the release of Honeycomb, with few enhancement requests being made in particular for calendar (8 requests), email (4 requests) and keyboard (9 requests) features. The numbers of requests were also low after the release of Ice Cream Sandwich, although contacts (19 requests), notification (10 requests), call (13 requests), calendar (11 requests) and text (16 requests) still featured in desired developments. Enhancement requests then surged after the release of Jellybean, for all of the features considered (contacts=83, screen=102, notification=97, call=53, calendar=37, email=58, text=50 and keyboard=62). Since the KitKat release there has been a reduction in calendar requests (only 4), and enhancement requests overall when compared to Jellybean. However, the Android community is still expressing their desire for further development in regard to a number of features, and most notably screen (41 requests), contacts (30 requests) and notifications (30 requests).

We performed formal statistical testing to see if the length of the Android OS release cycle and the project's maturity correlated with requests for enhancement. Our correlation test results confirmed that there were more requests between Android versions with longer release cycles (τ = 0.68, *P* < 0.05), and more requests for enhancement were logged as Android matured (τ = 0.67, *P* < 0.05). We also observed that significantly more enhancement requests were raised for Jellybean than for the other versions (*P* < 0.05 for all Mann-Whitney pairwise comparisons).

We then investigated how the numbers of followers varied over the lifetime of the OS. Of 61,528 followers altogether, Figure 4 (a) shows the largest percentage of community members were interested in Android's Early versions (correlation results supported our observation, τ = -0.64, *P* < 0.05). Enhancement requests for Cupcake also received a large degree of attention. Relatively less interest was shown in the issues raised in the middle Android versions, while Jellybean requests saw an upsurge in followers. In fact, in looking at the correspondence of followers and enhancement requests in Figure 4 (b) we observed that the early Android versions (from version 1.0 to 2.2) resulted in comparatively fewer enhancement requests when considered in light of the number of followers said to be interested in these requests, while the opposite result is evident for the latter versions (versions 2.3 to 4.4). This pattern of results was also seen when individual features (e.g., contacts, screen, notification, and so on) were considered.

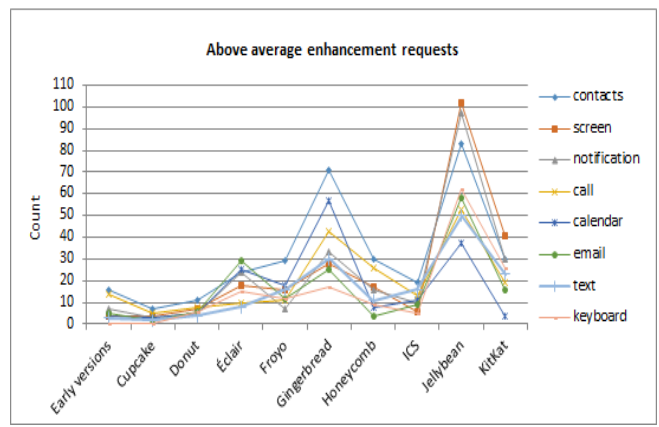

**Figure 3. Enhancement requests over Android releases**

## Enhancement Requests That Take Issue with Multiple Features (H3)

To test H3 we first conducted Kendall's tau-b correlation tests to examine how specific enhancement requests were related over the lifetime of the Android OS. These results, provided in Table 2, show that end-users expressed a strong desire for enhancements of some specific features together throughout the lifetime of the OS. For instance, Table 2 shows that when there were many requests for improvements to the contacts feature, those related to screen, notification, text and keyboard were raised with similarly higher frequency. The same is seen for screen and text and keyboard; notification and call, text and keyboard; call and text; email and text and keyboard; and text and keyboard. These results were strong ($\tau > 0.5$), statistically significant ($P < 0.05$) correlations.

Our PMI results for the top eight co-occurring enhancement requests, provided in Table 3, also reveal that end-users requested enhancements to contacts and search, screen and volume, and email and notification the most. In addition, screen, notification and contacts features were mentioned as candidates for enhancement with other features much more regularly than the other top features.

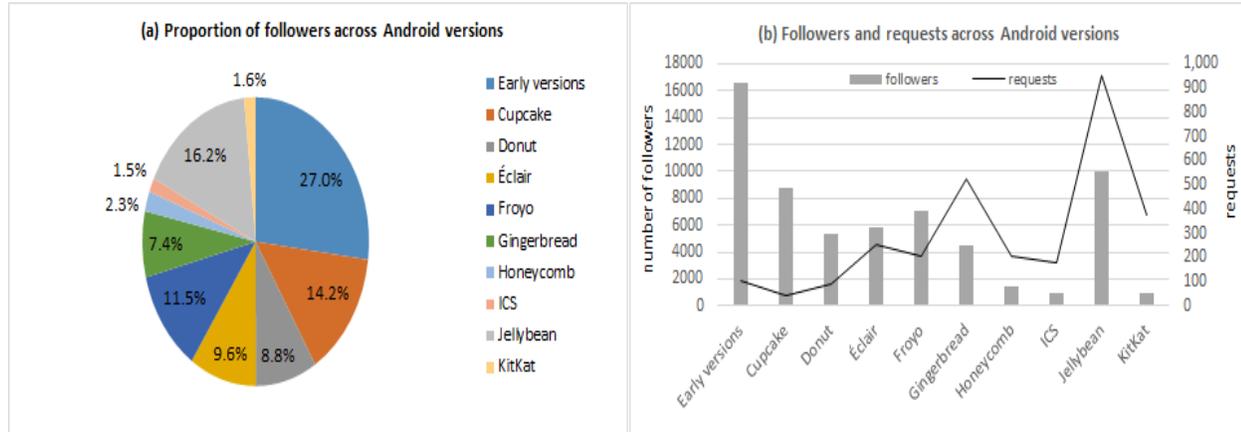

**Figure 4. Proportions and numbers of followers across Android versions and associated requests**

| Feature | 1 | 2 | 3 | 4 | 5 | 6 | 7 | 8 |
|---|---|---|---|---|---|---|---|---|
| 1 contacts | 1.0 | 0.75** | 0.84** | 0.76** | 0.53* | 0.58* | 0.80** | 0.71** |
| 2 screen |  | 1.0 | 0.79* | 0.49* | 0.52* | 0.63* | 0.66** | 0.96** |
| 3 notification |  |  | 1.0 | 0.72** | 0.52* | 0.63* | 0.71** | 0.75** |
| 4 call |  |  |  | 1.0 | 0.31 | 0.33 | 0.67** | 0.45 |
| 5 calendar |  |  |  |  | 1.0 | 0.63* | 0.57* | 0.56* |
| 6 email |  |  |  |  |  | 1.0 | 0.67** | 0.67** |
| 7 text |  |  |  |  |  |  | 1.0 | 0.71** |
| 8 keyboard |  |  |  |  |  |  |  | 1.0 |

Note: * strong statistically significant + correlation ($p < 0.05$), ** strong statistically significant + correlation ($p < 0.01$)

**Table 2. Correlations ($\tau$) among requests for enhancements**

| Feature A | Feature B | PMI |
|---|---|---|
| contacts | search | 2.04 |
| screen | volume | 1.36 |
| email | notification | 1.33 |
| call | screen | 0.89 |
| notification | screen | 0.57 |
| keyboard | screen | 0.47 |
| calendar | contacts | 0.44 |
| text | voice | 0.26 |

**Table 3. PMI for top enhancement requests**

## 5. Discussion and Implications

With a shift in paradigm towards frequent releases of system functionality to virtual communities, an increasing proportion of software is distributed via online application distribution platforms (OADPs) and then generally perfected after post-release usage (Ko et al. 2011), as against conventional methods where users are involved at project conception and throughout the development lifecycle (Gould and Lewis 1985). Thus, user reviews are key to the enhancement of such systems (Ko et al. 2011). We demonstrate this utility here, using: a) Pareto principle and power-law behavior (Boehm 1987), b) works examining the effects of the length of release cycles and project maturity on requests for enhancement (Hindle et al. 2011; Khomh et al. 2012), and c) complex network theory (Estrada 2011) as a basis for studying the widely used Android OS. We now revisit our findings in testing our three hypotheses (H1-H3), and consider our outcomes in relation to previous works.

### H1. Specific Android features are most often the subjects of enhancement requests.

Given our findings in the previous section, where the Android community identified features from contacts, screen, notification, call, calendar, email, text, and keyboard as being most frequently in need of enhancement, we must accept H1. That said, our outcomes did not conform to the 80%-20% rule previously observed by others (Boehm 1987; Louridas et al. 2008). Of note is that frequently requested features are all end-user facing, and thus likely affected a large cohort of users. Our findings here lend support to those previously noted by Kumar Maji et al. (2010), who found that app-related defects dominated the early versions of Android. Here we see a similar pattern in end-user calls for system enhancements. Although less noticeable, we also observed that the community was seeking enhancements to the Android OS "api". Such features would typically reside in the application frameworks and libraries, indicating that those developing apps for the Android OS also employed the issue tracker to express their requests for change. This highlights the particular opportunity afforded by OADPs where *a range* of user types are able to express their concerns in an effort to improve and extend previously deployed software. In fact, we observed that the attention of followers was primarily given to api, calendar, file, text, voice, and notification requests, and features with higher levels of enhancement requests also received greater levels of attention. This latter finding is fitting, and shows that the community's interest is appropriately focused.

### H2. Specific Android releases will have more requests for enhancement.

We observed that there were indeed more requests between Android versions with longer release cycles; more requests were logged as Android matured; and more requests were raised for Jellybean than for the other versions, we thus accept H2. These findings are in contrast to those uncovered by earlier work (Khomh et al. 2012). Our findings also contradict those reported earlier which established that the need for enhancements decreases as a software project matures (Hindle et al. 2011). We observed that more interest was shown in the enhancement requests logged for early versions, when there were far lower numbers of requests being made. In recent times, however, there have been increases in the numbers of both enhancement requests and followers. The goal of Google, and others making their products available via OADPs, should be to address end-users' enhancement requests in a timely manner, which could be informed by NLP techniques such as those used here.

### H3. Enhancement requests are likely to take issue with multiple features.

End-users logged issues about certain features together over Android's lifetime; and so again, we accept H3. For instance, requests for improvements to features for contacts, screen, notification, text and keyboard tended to occur together. Perhaps fixes for one aspect may address requests regarding the other? It may be inferred that the notification feature in the contacts app needs enhancement, which could be easily ascertained by deeper examination of those specific issues. However, the appearance of 'screen and volume' in the same review (noted in the previous section) provides a potential challenge to those charged with scheduling remedial work to app features. Perhaps a screen option to increase volume features is being sought by users? Given our ranking of these features, we are able to clearly identify which of the two features is in greater need of enhancement, but we would again need to employ deeper analysis approaches to gain additional context concerning user needs. In fact, identifying end-users' needs where multiple issues are reported is readily afforded given our identification of those specific issues where this duality occurs, confirming the potential utility of complex network theory (Estrada 2011). However, understanding *the nature* of their relationship(s) offers an avenue for future research.

## 6. Threats to Validity

Although the Android issue tracker is publicly hosted, and so is likely to capture a wide range of the community's concerns (Kumar Maji et al. 2010), issues may also be informally communicated and addressed within the development teams at Google. We also accept that there is a possibility that we could have missed misspelt features. Finally, we separated the issues based on the dates of the major Android OS releases, but accept that there is a possibility that some issues reported between specific releases were in fact related to earlier releases.